\documentclass[journal]{IEEEtran}
\usepackage{multirow}
\usepackage{cite}
\usepackage{amsmath,amssymb,amsfonts}
\usepackage{algorithmic}
\usepackage{graphicx}
\usepackage{comment}
\usepackage{dsfont}
\usepackage{bm}
\usepackage{stfloats}
\usepackage{pifont}
\usepackage{textcomp}
\usepackage{booktabs}
\usepackage{array}
\newcommand{\cmark}{\ding{51}}
\newcommand{\xmark}{\ding{55}}
\newenvironment{frcseries}{\fontfamily{frc}\selectfont}{}
\newcommand{\textfrc}[1]{{\frcseries#1}}
\def\BibTeX{{\rm B\kern-.05em{\sc i\kern-.025em b}\kern-.08em
    T\kern-.1667em\lower.7ex\hbox{E}\kern-.125emX}}
\begin{document}
\title{Co-VeGAN: Complex-Valued Generative Adversarial Network for Compressive Sensing\\ MR Image Reconstruction}
\author{Bhavya Vasudeva*, Puneesh Deora*, Saumik Bhattacharya, Pyari Mohan Pradhan
\thanks{* indicates equal contribution.}
\thanks{B. Vasudeva, P. Deora and P. M. Pradhan are with the Department of Electronics and Communication Engineering, IIT Roorkee, India (e-mail: bvasudeva@ec.iitr.ac.in; pdeora@ec.iitr.ac.in; pmpradhan.fec@iitr.ac.in).}
\thanks{S. Bhattacharya is with the Department of Electronics and Electrical Communication Engineering, IIT Kharagpur, India (e-mail: saumik@ece.iitkgp.ac.in).}
}

\maketitle

\begin{abstract}
Compressive sensing (CS) is widely used to reduce the acquisition time of magnetic resonance imaging (MRI). Although state-of-the-art deep learning based methods have been able to obtain fast, high-quality reconstruction of CS-MR images, their main drawback is that they treat complex-valued MRI data as real-valued entities. Most methods either extract the magnitude from the complex-valued entities or concatenate them as two real-valued channels. In both the cases, the phase content, which links the real and imaginary parts of the complex-valued entities, is discarded. In order to address the fundamental problem of real-valued deep networks, i.e. their inability to process complex-valued data, we propose a novel framework based on a complex-valued generative adversarial network (Co-VeGAN). Our model can process complex-valued input, which enables it to perform high-quality reconstruction of the CS-MR images. Further, considering that phase is a crucial component of complex-valued entities, we propose a novel complex-valued activation function, which is sensitive to the phase of the input. Extensive evaluation of the proposed approach on different datasets using various sampling masks demonstrates that the proposed model significantly outperforms the existing CS-MRI reconstruction techniques in terms of peak signal-to-noise ratio as well as structural similarity index. Further, it uses significantly fewer trainable parameters to do so, as compared to the real-valued deep learning based methods.
\end{abstract}


\section{Introduction}
Magnetic resonance imaging (MRI) is a frequently used medical imaging modality in clinical practice as it proves to be an excellent non-invasive source for revealing structural as well as anatomical information. A major shortcoming of the MRI acquisition process is its considerably long scan time. This is  due to sequential acquisition of large volumes of data, not in the image domain but in $k$-space, i.e. Fourier domain. Such a prolonged scanning time can cause significant artefacts because of physiological motion and movement of patient during the scan. It may also hinder the use of MRI in time-critical diagnosis. A possible way to speed-up the imaging process is 
leveraging compressive sensing (CS) \cite{donoho} based undersampling, which is used very often for this purpose. However, doing this renders the inverse problem ill-posed, making the recovery of high-quality MR images extremely challenging. 
\par Conventional approaches for CS-MRI reconstruction focus extensively on the usage of sparse representations to assume prior knowledge on the structure of the MR image to be reconstructed. Sparse representations can be explored by the use of predefined transforms \cite{lustig_sparse} such as total variation \cite{tv-csmri}, discrete wavelet transform \cite{dwt-csmri}, etc. Alternatively, dictionary learning based methods \cite{dlmri} learn sparse representations from the subspace spanned by the data. Both these types of approaches suffer from long computation time due to the iterative nature of the optimization processes. Moreover, the universally applicable sparsifying transforms might find it difficult to completely capture the fine details as observed in biological tissues \cite{sparsenotvalid}.
\par Deep learning based frameworks have enjoyed great success in similar inverse problems such as single-image super resolution \cite{sr}, denoising \cite{denoising}, etc., where the methods try to recover missing information from incomplete or noisy data. These successes as well as the advantages deep learning based frameworks offer over conventional methods, have prompted their use for CS-MRI reconstruction. The advantages worth mentioning are faster inference time and avoidance of explicit assumption of sparsity. 
Yang \textit{et al.} \cite{deepadmm} used the alternating direction method of multipliers (ADMM) algorithm \cite{admm} to train their deep network for CS-MRI reconstruction. With the recent advancements of generative adversarial networks (GANs) \cite{gan,patchgan}, CS-MRI reconstruction problems have also been addressed using adversarial learning frameworks. In a recent study, Bora \textit{et al.} \cite{bora} have shown that pretrained generative models like varitional autoencoders and GANs \cite{gan} can be used for recovery of CS signal without assuming sparsity. Yang \textit{et al.} \cite{dagan} proposed a U-net \cite{ronneberger2015u} based generator, following a refinement learning based approach, with mean squared error (MSE) and perceptual loss to reconstruct the images. The authors of \cite{refinegan} proposed a fully residual network using addition-based skip connections. They used cyclic loss for data consistency constraints in the training process to achieve better reconstruction quality. Deora \textit{et al.} \cite{robustgan} proposed a U-net based generator with a patch-based discriminator \cite{patchgan} to perform the reconstruction task. Along with mean absolute error (MAE) and structural similarity (SSIM) \cite{ssim}, authors used Wasserstein loss \cite{wgan} to improve the adversarial learning. Mardani \textit{et al.} \cite{gancs} introduced affine projection operator in-between the generator and the discriminator to improve the data consistency in the reconstructed images.
\begin{figure}[h!]
    \centering 
    \includegraphics[scale=0.36]{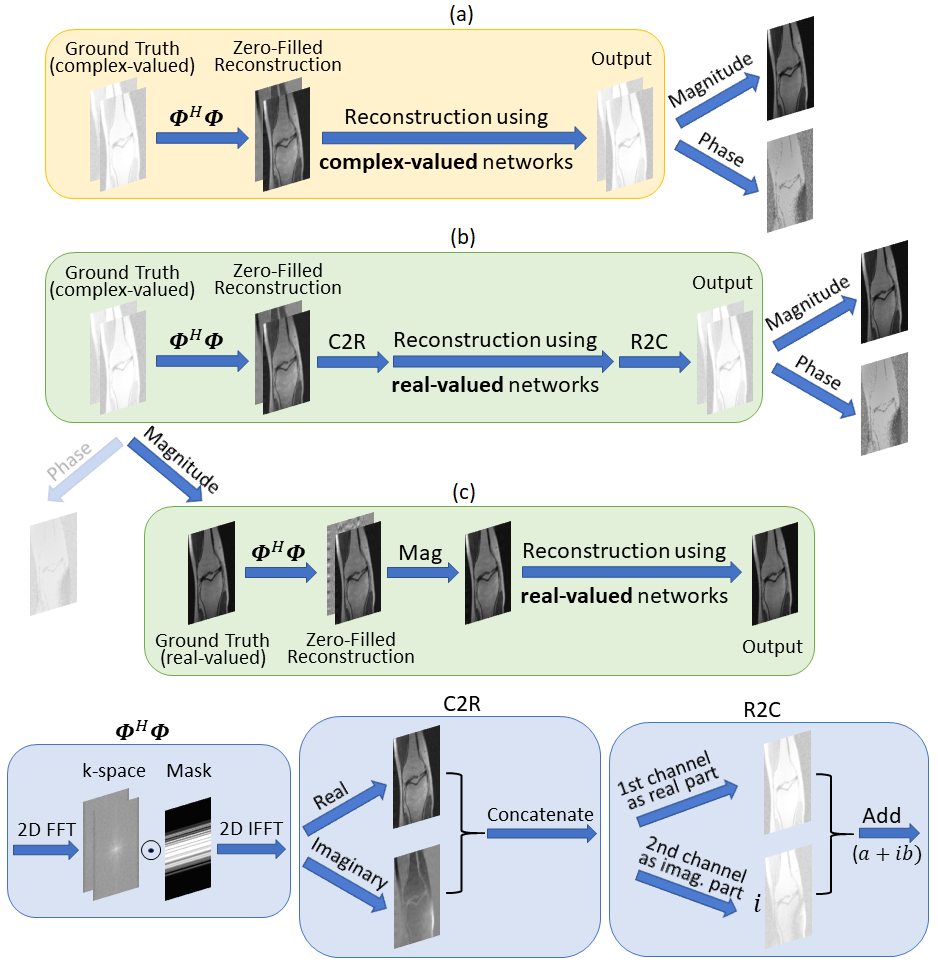}
    \caption{Comparison of the proposed pipeline (a) with the pipelines (b), (c) followed by SOTA deep learning based methods for CS-MRI reconstruction. Our method uses complex-valued operations at all stages to process the complex-valued MRI data, preserving both the magnitude and phase content. In contrast, the other methods treat the complex-valued MRI data as real-valued data by either concatenating the real and imaginary parts as two real-valued channels or by using the magnitude images. $\Phi$ denotes the matrix for undersampling in the $k$-space, and $\Phi^H\Phi$ denotes the matrix to obtain the zero-filled reconstruction (ZFR). The two channels shown in ground truth, ZFR, $k$-space, output represent the complex-valued data.}
    \label{fig1}
\end{figure}

\par Although state-of-the-art (SOTA) deep learning based approaches have significantly improved the quality of reconstruction as well as the inference time as compared to the conventional iterative approaches, they treat the complex-valued MRI data as real-valued data. This is done either by splitting and concatenating the real and imaginary parts of the MRI data as two real-valued channels \cite{variational, gancs, refinegan, modl}, as illustrated in Fig. \ref{fig1}(b), or by using the magnitude only images \cite{fbpcnet,Hyun}, as shown in Fig. \ref{fig1}(c). Deep neural networks with real-valued weights are then used to process such data. The use of the latter pipeline leads to reconstructed outputs providing only magnitude images and losing out on phase information. Other works \cite{variational, gancs, refinegan, modl}, where reconstructed outputs provide phase information in addition to the magnitude, do so by processing the real and imaginary components of MRI using real-valued filters. Such an approach essentially decouples the real and imaginary parts of the complex-valued MRI data, thus destroying the phase information of MRI when it is processed as well as leading to unsatisfactory recovery of phase information. There are several applications which utilize the phase content present in the complex-valued data, such as phase-contrast imaging, quantifying velocity, and measuring blood flow and volume flow \cite{mri_app}, which is why it is crucial to faithfully recover the phase content. 

\par Moreover, when the dataset itself contains only magnitude images, the aforementioned methods and others \cite{dagan, robustgan} use a pipeline similar to the one shown in Fig. \ref{fig1}(c). However, since the zero-filled reconstruction (ZFR) is complex-valued in nature, processing it with real-valued networks fails to utilize the complex algebraic structure of the data. 
\par The inability of real-valued deep learning based methods to process complex-valued data is a fundamental weakness which causes issues in all the aforementioned cases. Leveraging complex-valued neural networks to process the inherently complex-valued MRI data (Fig. \ref{fig1}(a)) naturally appears to be the right direction to proceed.

\par The motivation for using complex-valued networks stems from the benefits provided by complex parameter space as compared to real parameter space. Apart from having biological inspiration and significance \cite{biologicalmotiv}, complex-valued representation not only increases the representational capacity of the network, it can be more stable than real-valued space in various applications \cite{deepcomplex,danihelka}. Complex-valued operations can be performed with easy optimization techniques \cite{nitta} without sacrificing the generalization ability of the network \cite{hirose2012generalization}. Several researchers have reported that the complex-valued neural network exhibits faster learning with better robustness to noise \cite{danihelka,arjovsky,wisdom2016full}. Complex-valued operations also preserve the phase information which encodes fine structural details of an image \cite{spmotiv2,deepcomplex}. The importance of phase in images in terms of recovery of magnitude information has been discussed in \cite{spmotiv2}. A more mathematical motivation for the use of complex weights in convolutional neural networks is presented in \cite{brunamathematical}. Even with these striking benefits, complex-valued deep networks are not widely explored. Only a few works have explored their use in applications such as vision \cite{deepcomplex}, \cite{cnnpolsar}, parallel MR imaging \cite{parallelwang}, MRI fingerprinting \cite{cardiod}, audio-related tasks \cite{deepcomplex}. 
\par We seek to exploit the benefits of complex-valued neural networks in order to overcome the problems suffered by the SOTA deep learning based CS-MRI reconstruction frameworks. Considering the tremendous capabilities of adversarial learning approaches in MRI reconstruction, as mentioned earlier, we propose a novel complex-valued GAN (Co-VeGAN) which, to the best of our knowledge, has never been explored for a reconstruction problem.
\par Activation functions in deep neural networks facilitate the networks to learn highly complex functions due to their nonlinear nature. Real-valued activation functions in neural networks have been extensively studied in literature, with sigmoid, rectified linear unit (ReLU), leaky ReLU, parametric ReLU \cite{prelu} etc. being widely used. However, these activation functions do not prove to be as effective when their complex-valued equivalents are considered for complex-valued networks. This can be attributed to their weak or limited sensitivity to the changes in input phase. Such a behaviour has been studied by Virtue \textit{et al.} \cite{cardiod} for MRI fingerprinting, where they propose a cardioid activation function, which is not only sensitive to input phase but also preserves it at the output. We argue that such phase preservation can compromise with the flexibility of the activation function. Moreover, cardioid has a non-trainable, fixed profile which does not provide room for adaptation and causes it to favour certain types of inputs. To overcome such limitations, we formulate a novel activation function by leveraging weighted sinusoid functions. We further demonstrate the usefulness of the components of the proposed activation by studying its variants, and show that cardioid turns out as a specific case.

\par Another limitation pertaining to the current frameworks is the widespread use of pixel-wise $l_2$ or $l_1$ loss functions for training. This efficiently reconstructs the low-frequency components of the image, but often fails to generate the mid and high-frequency information which depicts fine textural and structural parts of an image \cite{wavelet_srnet}. We introduce a novel Gaussian-weighted wavelet loss function, which helps to resolve this issue and focuses on reconstruction of mid-frequency components of the MR image.
\par We believe that this work will provide the basis for future works deploying complex-valued GAN based frameworks not only for CS-MRI reconstruction, but also for other medical imaging applications, which involve the use of complex-valued data, such as MRI super-resolution, denoising, segmentation, etc. 
The major contributions of this work are summarized below:
\begin{itemize}

\item We propose a novel Co-VeGAN framework and explore it for CS-MRI reconstruction. In contrast to the SOTA deep learning based frameworks for CS-MRI reconstruction, which essentially consider the inherently complex-valued MRI as real-valued data, our framework takes advantage of its complex algebraic structure by using complex-valued operations. Through extensive experiments, we show that doing so not only provides superior reconstruction of magnitude information but also leads to accurate recovery of phase content, which is either completely lost or unsatisfactorily recovered in the existing studies.

\item We propose a novel activation function designed specifically for complex-valued entities. Our activation function takes into account the limitations of the existing activation functions - poor transferability of real-valued activations to the complex domain, as well as limited sensitivity to input phase and fixed profile of cardioid activation.
\item We introduce a novel variant of wavelet loss function to overcome the limitations posed by $l_2$ or $l_1$ norm based losses in reconstruction problems. 
\item In addition to this, we focus on the architectural aspects of the complex deep neural networks used in our study. Specifically, we propose a novel dense U-net architecture leveraging dense connections within the contracting and expanding paths to allow feature reuse between layers and improve the information flow.
\item We perform comprehensive ablation studies to show the significance of each proposed element by analyzing various network settings as well as various activation functions. We also analyze the performance of equivalent real-valued networks, and show that our framework not only provides superior quality reconstructions results, but also uses significantly fewer trainable parameters, thus minimizing the storage requirements. 
\item We test our proposed approach on three different datasets (both real and complex-valued) and show that it outperforms both conventional and SOTA deep learning frameworks for CS-MRI reconstruction. 

\end{itemize}

\section{Methodology}
The acquisition model of CS-MRI can be described as follows:
\begin{equation}
\mathbf{y}  = \mathbf{\Phi} \mathbf{x} + \bm{\zeta} = \mathbf{U}\mathbf{F} \mathbf{x} + \bm{\zeta},
\end{equation}
where $\mathbf{x} \in \mathbb{C}^{K^2}$ is the desired image in vector form, $\mathbf{y} \in \mathbb{C}^{M}$ denotes the observed data vector, the vector $\bm{\zeta} \in \mathbb{C}^{M}$ captures the noise. $\mathbf{F} \in \mathbb{C}^{K^2 \times K^2}$ denotes the matrix to compute the 2D Fourier transform, $\mathbf{U} \in \mathbb{R}^{M \times K^2}$ describes the matrix for undersampling. Given an observation $\mathbf{y}$, the aim of reconstruction is to recover $\mathbf{x}$ in the presence of a non-zero $\bm{\zeta}$. 
\par We attempt the recovery of $\mathbf{x}$ by using a GAN model. A GAN comprises of two networks, namely a generator and a discriminator. In order to generate images which are similar to the samples of the distribution of true data $\mathbf{y}_{t}$, the generator $G$ attempts to map an input vector $\mathbf{z}$ to the output $G(\mathbf{z})$. On the other hand, the aim of the discriminator $D$ is to classify the generated samples $G(\mathbf{z})$ and the samples from the distribution of $\mathbf{y}_{t}$. 
\par As mentioned earlier, for the problem of CS-MRI reconstruction, both the observation $\mathbf{y}$ and the desired image $\mathbf{x}$ are complex-valued, whereas the parameters of a GAN are real-valued in existing methods. Thus, we propose a Co-VeGAN framework, which is described below.
\vspace{-2mm}
\subsection{Complex-valued GAN}
\par We propose a complex-valued GAN which consists of a complex-valued $G: \mathbb{C}^{K \times K} \rightarrow\mathbb{C}^{K \times K}$, and a real-valued $D: \mathbb{R}^{K \times K} \rightarrow\mathbb{R}^{1 \times 1}$. Although the parameters of both $G$ and $D$ can be complex-valued, we opt for the use of a real-valued discriminator, mainly because doing so enables it to take the visual soundness of the magnitude images into account while discriminating the generated and actual images. The use of a real-valued discriminator also makes our framework suitable for real-valued datasets, where the ground truth (GT) is real-valued. In order to constrain the phase content of the generated output to be similar to that of the GT, we use a pixel-wise loss, described in section II-D.  
\vspace{-2mm}
\subsection{Complex-valued operations}
\par In this section, we discuss the important operations required for the implementation of a complex-valued network, namely convolution, backpropagation, batch normalization (BN) \cite{bn}, and activation. 
\subsubsection{Convolution} The complex-valued equivalent of real-valued 2D convolution is discussed below. The convolution of a complex-valued kernel $\mathbf{W} = \mathbf{W_R} + \textfrc{i} \mathbf{W_I}$  with complex-valued feature maps $\mathbf{F} = \mathbf{F_R} + \textfrc{i}\mathbf{F_I}$, can be represented as $\mathbf{A} = \mathbf{W*F} =\mathbf{A_R} + \textfrc{i}\mathbf{A_I}$, where $*$ denotes convolution operation, $\textfrc{i}$ denotes $\sqrt{-1}$, and
\begin{equation}
\begin{split}
 \mathbf{A_R} & =  \mathbf{W_R*F_R - W_I*F_I}, \\
         \mathbf{A_I} & =\mathbf{W_R*F_I + W_I*F_R}
         \label{conv},
\end{split}
\end{equation}
similar to complex-valued multiplication. In these notations, the subscripts $\mathbf{R}$ and $\mathbf{I}$ denote the real and imaginary parts of the complex-valued entities, respectively. In order to implement the aforementioned complex-valued convolution, we make use of real-valued tensors, where $\mathbf{W}$ ($\mathbf{F}$) is stored such that the imaginary part $\mathbf{W_I}$ ($\mathbf{F_I}$) is concatenated to the real part $\mathbf{W_R}$ ($\mathbf{F_R}$). The resultant includes four real-valued 2D convolutions as defined in (\ref{conv}), and is stored in a similar manner by concatenating $\mathbf{A_I}$ to $\mathbf{A_R}$.
\subsubsection{Backpropagation} Backpropagation can be performed on a function $f$ that is non-holomorphic as long as it is differentiable with respect to its real and imaginary parts \cite{cderivative}. Since all the loss functions considered in this work are real-valued, we consider $f$ to be a real-valued function of $l$-dimensional weight vector $\mathbf{w}$. The update rule of $\mathbf{w}$ using gradient descent is written as:
\begin{equation}
    \mathbf{w} = \mathbf{w} - \rho \nabla_{\bar{\mathbf{w}}}f(\mathbf{w}),
\end{equation}
 where $\rho$ is the learning rate (LR), $\bar{\mathbf{w}}$ denotes the complex conjugate of $\mathbf{w}$, and the gradient of $f$ is calculated as:
\begin{equation}
\begin{split}
 \nabla_{\bar{\mathbf{w}}}f(\mathbf{w}) &= \Big[ \frac{\partial f}{\partial \bar{w_1}} \cdots \frac{\partial f}{\partial \bar{w_l}}\Big ]^T\\
 \frac{\partial f}{\partial \bar{w}} &= \frac{1}{2} \Big( \frac{\partial f}{\partial w_R} +\textfrc{i} \frac{\partial f}{\partial w_I}\Big ).
\end{split}
\end{equation}
\subsubsection{Batch Normalization} We make use of the complex BN (CBN) applicable to complex numbers, proposed in \cite{deepcomplex}. To ensure that the complex data is scaled in such a way that the distribution of real and imaginary components is circular, the 2D complex vector can be whitened as shown below:
\begin{equation}
 \mathbf{x}_{std} = \mathbf{B}^{-\frac{1}{2}}(\mathbf{x} - \mathbb{E}[\mathbf{x}]), 
\end{equation}
where $\mathbf{B}$ denotes the covariance matrix, and $\mathbb{E}$ denotes expectation operator. $\mathbf{B}$ can be represented as:
\begin{equation}
 \mathbf{B} = \begin{bmatrix}
 \mathrm{Cov}(\mathbf{x_R},\mathbf{x_R}) & \mathrm{Cov} (\mathbf{x_R}, \mathbf{x_I}) \\
 \mathrm{Cov}(\mathbf{x_I},\mathbf{x_R}) & \mathrm{Cov} (\mathbf{x_I}, \mathbf{x_I})
 \end{bmatrix}.
\end{equation}
 Learnable parameters $\bm{\gamma}$, $\bm{\beta}$ are used to scale and shift the aforementioned standardized vector as follows:
\begin{equation}
    \mathbf{x}_{BN} = \bm{\gamma}\mathbf{x}_{std} + \bm{\beta},    
\end{equation}
where \bm{$\gamma$} is a $2\times2$ matrix, and $\bm{\beta}$ is a complex number.

\subsubsection{Activation} In order to work with complex-valued entities, several activation functions have been proposed in previous works \cite{arjovsky}. One common class of complex activation functions is obtained when the same nonlinearity is applied separately on the real and imaginary parts. The complex equivalent of the commonly used ReLU ($\mathbb{C}$ReLU) is given by:
    \begin{equation}
    \mathbb{C}\text{\textit{ReLU}}(a)=ReLU(a_R)+\textfrc{i}ReLU(a_I),
    \end{equation}
where $a$ is a complex-valued input. Similarly, the expression for the complex equivalent of parametric ReLU \cite{prelu} ($\mathbb{C}$PReLU) is formulated as follows:
\begin{equation*}
\mathbb{C}\text{\textit{PReLU}}(a)=  \begin{cases} 
 a_R+\textfrc{i}a_I & 0\leq a_R, 0\leq a_I, \\
 \beta_Ra_R+\textfrc{i}a_I & a_R< 0, 0\leq a_I, \\
 a_R+\textfrc{i}\beta_Ia_I & 0\leq a_R, a_I< 0, \\
 \beta_Ra_R+\textfrc{i}\beta_Ia_I & a_R<0, a_I<0,
  \end{cases}
\end{equation*}
where $\beta_R$ and $\beta_I$ are trainable parameters. In this way, the complex equivalent of any activation function which works well for real-valued networks can be explored for complex-valued networks. 
\par There is another class of complex activation functions where the nonlinearity is applied on the magnitude $|a|$ and phase $\angle a$ of $a$. One such activation is $z$ReLU \cite{zrelu}, which allows a complex element to pass only if it lies in the first quadrant. It is given by:
     \begin{equation}
    z\text{\textit{ReLU}}(a)=  \begin{cases} 
      a & 0\leq\angle a\leq\frac{\pi}{2}, \\
      0 & otherwise.
   \end{cases}
    \end{equation}
\par However, all the aforementioned activation functions are weakly sensitive to the changes in the phase of the input, which is not only important from the biological perspective but also crucial for the complex representation. 
Virtue \textit{et al.} \cite{cardiod} have proposed the cardioid activation, where the input phase is preserved at the output, and the magnitude of the output is sensitive to the phase of the input. It is formulated as:
\begin{equation}
    \text{\textit{Cardioid}}(a)=\text{\textit{g}}(\angle a)a=\frac{1}{2}(1+cos(\angle a))a,
    \label{card_eq}
\end{equation}
where \textit{g}$(\cdot)$ denotes the gain, which is positive-valued for phase preservation. 

\par However, phase preservation reduces the flexibility of the activation function because change in the phase from the input to the output can be viewed as the delay of a neuron. Moreover, since the phase is modified at the CBN layer, its preservation at the activation layer might not prove useful. Another drawback of cardioid is that the gain is fixed, and its shape is not learnable. Due to its fixed shape, it favours positive real-valued inputs over negative ones, as \textit{g}$(0)=1$ and \textit{g}$(\pi)=0$. Also, it is designed to favour real-valued inputs over purely imaginary inputs, as \textit{g}$(\frac{\pi}{2})=0.5$, without much intuition. 
\begin{figure*}[t]
    \centering 
    \includegraphics[scale=0.4]{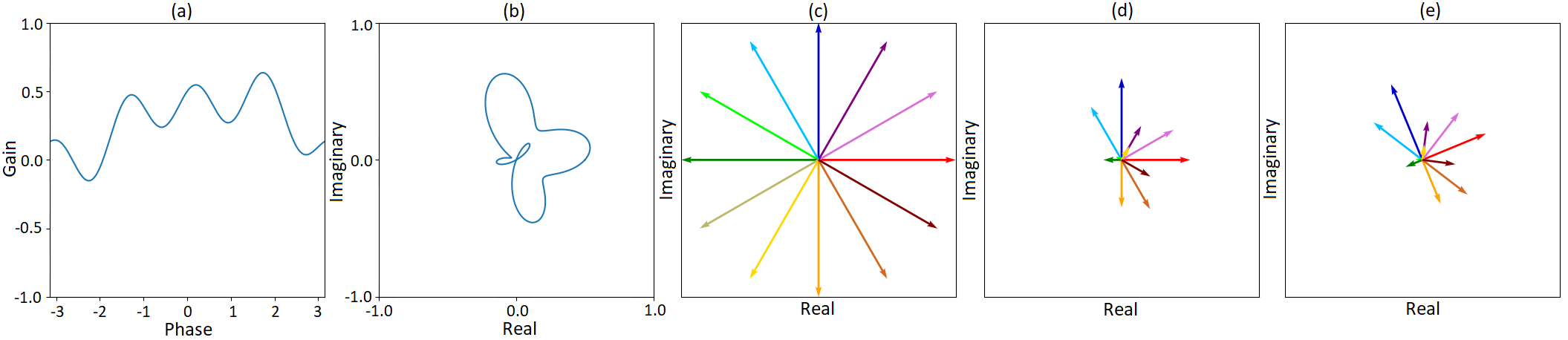}
    \caption{An example of the proposed activation PC-SS, with the set of values: [$w_0$, $w_1$, $w_2$, $\theta_0$, $\theta_1$, $\theta_2$] $=$ [0.08, -0.04, 0.06, 0.6, 0.4, 0.2]. (a) Plot of gain vs input phase, (b) plot of gain in complex space, (c) a set of complex-valued inputs, (d) outputs of PC-SS using gain shown in (a) with $\phi$ = 0, and (e) outputs with $\phi$ = $\frac{\pi}{8}$.}
    \label{act_plot}
\end{figure*}
\par We propose a novel activation function, where each neuron can introduce a phase change (PC) $\phi$ in the input. This can also be viewed as allowing coupling between $a_R$ and $a_I$. In order to allow the magnitude of the output to be sensitive to the phase of the input, any continuous function of the input phase can be used as the gain. Also, this function should be periodic with period $2\pi$, so that it is continuous for all values of the phase. We propose the use of a weighted sum of sinusoids (SS) in this work, to provide flexibility and maintain periodicity. The proposed PC-SS is formulated as follows:
\begin{equation}
\label{pcsos}
    \text{\textit{PC-SS}}(a)=\frac{\sum\limits_{p=0}^{P_S-1}w_p\{1+cos\left(2^p(\angle a -\theta_p)\right)\}}{2\sum\limits_{p=0}^{P_S-1}|w_p|+\epsilon}ae^{i\phi},
\end{equation}
where $w_p$, $\theta_p$ and $\phi$ are trainable parameters, $P_S$ denotes the number of sinusoids, and $\epsilon$ is a small constant to avoid division by 0. The use of trainable parameters $w_p$ and $\theta_p$ allow the framework to learn suitable gain functions. The number of extra parameters introduced by PC-SS is very small as compared to the total number of weights already present in the framework, since for a particular channel of a layer, the activation shares its weights. The numerator is normalized by the sum of the absolute value of the weights, so that the gain remains bounded in $[-1,1]$, to avoid the problem of exploding gradient. As the gain can take negative values, the phase may not be preserved. In order to control the trade-off between the flexibility and the increase in the number of trainable parameters, $P_S$ is set as 3 in this study. Fig. \ref{act_plot} illustrates an example of PC-SS. It is evident that PC-SS has a very flexible learnable shape, and that it can cause changes in the phase of the complex inputs. 

\par In order to demonstrate the importance of allowing negative gain and using $\phi$ in PC-SS, we consider two of its variants. In the first case, $\phi$ is set as 0 and absolute value of weights is taken in the numerator as well, so that the gain is bounded in $[0,1]$. This is the phase preserving (PP) equivalent of PC-SS, and is named as PP-SS. When $w_0=1$ and the rest of the parameters are set as 0, PP-SS is reduced to cardioid. In order to highlight the importance of $\phi$, we consider another variant, where negative gain is allowed, as in (\ref{pcsos}), but $\phi$ is set as 0. In this case, the negative gain may change the phase, but the value of $tan^{-1}(\frac{a_I}{a_R})$ remains the same at the output. Thus, this tangent inverse preserving (TIP) equivalent of PC-SS is named as TIP-SS.
\vspace{-2mm}
\subsection{Network Architecture}
\par
\begin{figure*}[t]
    \centering
    \includegraphics[scale=0.58]{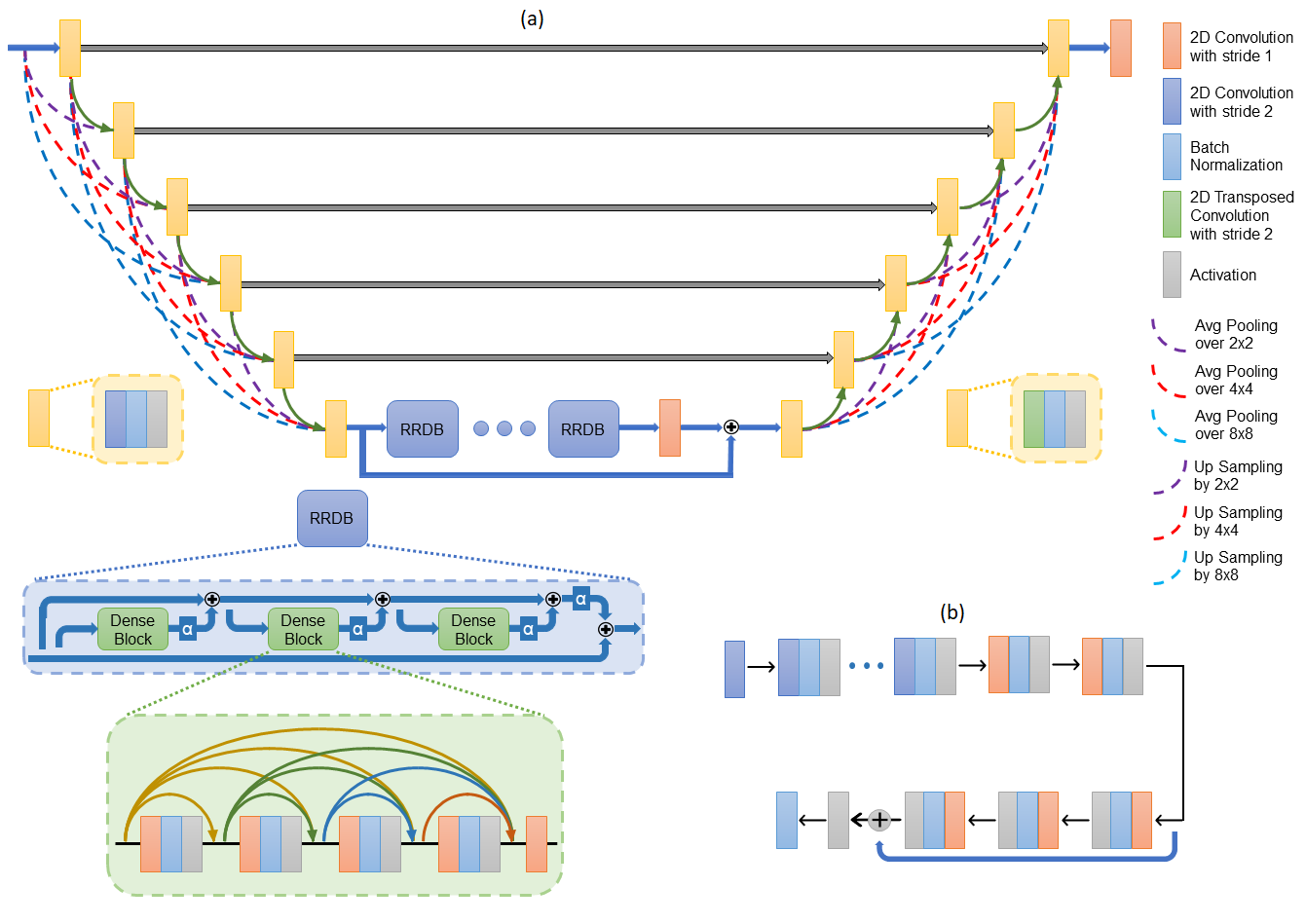}
    \caption{(a) Proposed dense U-net generator architecture with RRDBs and (b) discriminator architecture.}
    \label{gen}
\end{figure*}

 The generator architecture used in the proposed framework is shown in Fig. \ref{gen}(a). It is based on a U-net architecture. The left side is a contracting path, where each step involves the creation of downsampled feature maps using a convolutional layer with stride 2, which is followed by CBN and activation function. The right side is an expanding path, where each step consists of upsampling (by a factor of two), convolutional layer to create new feature maps, followed by CBN and activation layers. In order to provide richer context about the low-level features for superior reconstruction, the low-level feature maps from the contracting path are concatenated to the high-level feature maps of same size in the expanding path. 
\par We propose the use of dense connections between the steps (layers) within the contracting as well as the expanding path. These dense connections \cite{dense} help improve the flow of information between the layers and encourage the features from the preceding layers to be reused. There is an added benefit of increase in variation of available information by concatenation of feature maps. Since the feature maps at various layers are not of the same size, average pooling and upsampling (with bilinear interpolation) operations have been introduced in the dense connections between the layers of the contracting path and the expanding path, respectively. However, the use of these operations to change the size of the feature maps by a factor greater than $r$ (less than size $\frac{K}{2^r}\times \frac{K}{2^r}$) not only increases the computational and memory requirement, but also reduces the quality of information available to the subsequent layers. As shown in Fig. \ref{gen}(a), $r$ is set as 3 in this work.

\par Further, residual-in-residual dense blocks (RRDBs) \cite{rrdb} are incorporated at the lowest layer of the generator, where feature maps of size $\frac{K}{32}\times\frac{K}{32}$ are present. Each block uses residual learning across each dense block, as well as across a group of three dense blocks. At both the levels, residual scaling is used, i.e. the residuals are scaled by $\alpha$ before they are added to the identity mapping, as shown in Fig. \ref{gen}(a). These RRDBs not only make the network length variable, because of the residual connections which make identity mappings easier to learn, but also make a rich amount of information accessible to the deeper layers, through the dense connections. They also help to overcome the problem of vanishing gradients at the shallower layers. At the output of the generator, a hyperbolic tangent activation is applied.
\par The discriminator architecture is based on a standard convolutional neural network, as shown in Fig. \ref{gen}(b). It has 11 convolutional layers, each of which is followed by BN and leaky ReLU activation function. We use a patch-based discriminator to increase the focus on the reconstruction of high-frequency content. The patch-based discriminator scores each patch of the input image separately, and its output is the mean of the individual patch scores. This framework makes the network insensitive to the input size.  
\vspace{-3mm}
\subsection{Training Losses}
\subsubsection{Adversarial Loss} In order to constrain the generator to produce the MR image corresponding to the samples acquired in the $k$-space, it is conditioned \cite{cgan} over the ZFR given by:
\begin{equation}
\mathbf{x_u}  = \mathbf{\Phi}^H\mathbf{y} = \mathbf{F}^H\mathbf{U}^H\mathbf{y},
\end{equation}
where $\mathbf{x_u} \in \mathbb{C}^{K^2}$, $H$ denotes the Hermitian operator. Conventionally, the solution to the minmax game between the generator and the discriminator is obtained by using binary cross-entropy based loss. However, it causes the problem of vanishing and exploding gradients, which makes the training of the GAN model unstable. In order to prevent this, Wasserstein distance based loss \cite{wgan} is used. Mathematically, the training process of this conditional GAN using Wasserstein loss is formulated in the following way:
\begin{equation}
\begin{split}
    \min_{G}\max_{D}L_{GAN} & = \mathbb{E}_{\mathbf{x}\sim p_x(\mathbf{x})}[D(|\mathbf{x}|)]\\ & -\mathbb{E}_{\mathbf{x_u}\sim p_{x_u}(\mathbf{x_u})}[D(|G(\mathbf{x_u})|)],
    \end{split}
\end{equation}
where $p_x(\mathbf{x})$ is the distribution of the GT images, and $p_{x_u}(\mathbf{x_u})$ is the distribution of the aliased ZFR images.
\par In order to solve this optimization problem, an alternating process of updating $G$ once and $D$ $n_D$ times is followed. In order to enforce the Lipschitz constraint on $D$, weight clipping is applied on its weights \cite{wgan}.
\subsubsection{Content Loss} Besides adversarial loss, other losses are required to bring the reconstructed output closer to the corresponding GT image. In order to do so, we incorporate an MAE based loss, so that the pixel-wise difference between the GT and the generated image is minimized. It is given by:
\begin{equation}
    L_{\ell_1} = \mathbb{E}[\|G(\mathbf{x_u})-\mathbf{x}\|_1],
\end{equation}
where $\|\cdot\|_1$ denotes the $\ell_1$ norm. The $\ell_1$ norm is preferred over the $\ell_2$ norm which can lead to overly smooth and blurry reconstruction. In addition to this, enforcing $\ell_1$ similarity in the complex domain helps in accurate phase as well as magnitude reconstruction. 

\subsubsection{Structural Similarity Loss} As the  high-frequency details in the MR image help in distinguishing various regions with structural information, it is extremely important to improve their reconstruction. SSIM quantifies the similarity between the local patches of two images on the basis of luminance, contrast and structure. It is calculated as follows:
\begin{equation}
    SSIM(\textbf{M},\textbf{N})=\frac{2\mu_M\mu_N+\epsilon_1}{\mu_M^2+\mu_N^2+\epsilon_1}\frac{2\sigma_{MN}+\epsilon_2}{\sigma_M^2+\sigma_N^2+\epsilon_2},
\end{equation}
where $\mathbf{M}$ and $\mathbf{N}$ represent two image patches, $\mu_M$ and $\mu_N$ denote their means, $\sigma_M^2$ and $\sigma_N^2$ denote their variances, $\sigma_{MN}$ denotes their covariance, and $\epsilon_1$, $\epsilon_2$ are slack values to avoid division by zero. 
\par In order to improve the perceptual quality of the reconstructed MR image and preserve the structural details, a mean SSIM (mSSIM) based loss is incorporated in the training of the generator. It maximizes the patch-wise SSIM between the reconstructed output and the corresponding GT image by minimizing the following expression:
\begin{equation}
    L_{mSSIM} = 1-\mathbb{E}\left[\frac{1}{P}\sum_{p=1}^{P}SSIM(|G_p(\mathbf{x_u})|,|\mathbf{x}_p|)\right],
\end{equation}
where $P$ denotes the number of patches in the image. Since the SSIM is used to compare the perceptual similarity between two images, we use magnitude GT and reconstructed images in the aforementioned formulation. 

\subsubsection{Wavelet Loss} 
\begin{figure}[h!]
    \centering 
    \includegraphics[scale=0.35]{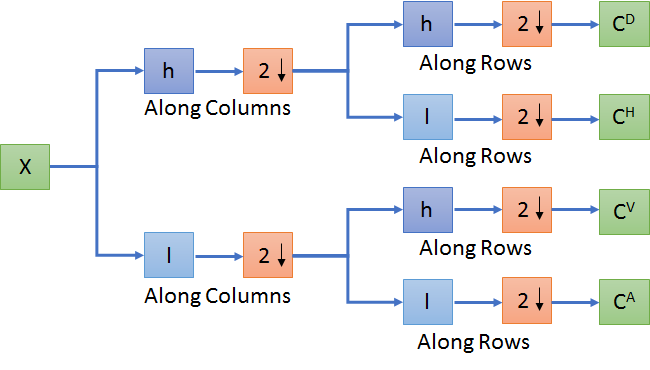}
    \caption{Illustration of the recursive process followed to compute the wavelet packet decomposition of $\textbf{X}$. In this work, Haar wavelet is used, so that $h=(\frac{1}{\sqrt{2}},\frac{1}{\sqrt{2}})$ and $l=(\frac{1}{\sqrt{2}},\frac{-1}{\sqrt{2}})$.}
    \label{wvt_fig}
\end{figure}

In order to further enhance the textural details in the generated image, a weighted version of MAE in the wavelet domain is introduced as another loss term. This is inspired by another inverse problem which is closely related to CS-MRI reconstruction, namely super-resolution \cite{wavelet_srnet}. In order to decompose the image into sets of wavelet coefficients $\mathbf{C}$, which are equal in size, and correspond to even division of bands in the frequency domain, the wavelet packet transform is used. Fig. \ref{wvt_fig} depicts one step of the recursive process which is followed to obtain the sets of wavelet coefficients. For an $r_w$ level decomposition which produces $P_w=4^{r_w}$ sets of wavelet coefficients of size $K_w\times K_w$ with $K_w=\frac{K}{\sqrt{P_w}}$, the wavelet loss is formulated as follows:
\begin{equation*}
    L_{wvt} = \frac{1}{P_wK_w^2}\sum_{p=1}^{P_w}\gamma_p\left [\sum_{i,j=1}^{K_w}|\mathbf{C}^p_{|G(\mathbf{x_u})|}(i,j)-\mathbf{C}^p_\mathbf{|x|}(i,j)|\right],
\end{equation*}
where $\gamma_p$ denotes the weight of the $p^{th}$ set of coefficients. Since the pixel-wise MAE loss contributes more towards the improvement of low-frequency details, and the mSSIM loss focuses more on preserving the high-frequency content in the reconstructed image, higher weights $\gamma_p$ are assigned to the wavelet coefficients corresponding to the band-pass components to improve their reconstruction. This is done by setting the weights according to the normalized probability density function of a Gaussian distribution with mean $\frac{(P_w-1)}{2}$ and variance $\sigma^2_w$. In this work, $r_w=3$, i.e. $P_w=64$, and $\sigma^2_w=12.5$. 
\begin{table*}[h!]
\centering
\large
\caption{Quantitative results for ablation study of the proposed model}
\label{tab_abl}
\resizebox{1.9\columnwidth}{!}{
\begin{tabular}{ccccccc} 
 \toprule
 \rule{0pt}{9pt} Network Settings & $1^{st}$ & $2^{nd}$ & $3^{rd}$ & $4^{th}$ & $5^{th}$ & $6^{th}$\\
 \midrule
\rule{0pt}{9pt} Complex-valued GAN & \xmark & \cmark & \cmark & \cmark & \cmark & \xmark \\ 
\rule{0pt}{9pt} RRDBs & \xmark & \xmark & \cmark & \cmark & \cmark & \cmark \\
\rule{0pt}{9pt} Dense U-net & \xmark & \xmark & \xmark & \cmark & \cmark & \cmark \\
\rule{0pt}{9pt} Wavelet loss & \xmark & \xmark & \xmark & \xmark & \cmark & \cmark \\
\midrule
\midrule
 \rule{0pt}{6pt}Generator parameters & 2M & 1.2M & 1.5M & 1.7M & 1.7M & 3.5M\\
 \midrule
 \rule{0pt}{9pt} PSNR (dB) / mSSIM & 39.640 / 0.9823 & 40.048 / 0.9866 & 41.418 / 0.9879 & 43.798 / 0.9902 & \textbf{45.044} / \textbf{0.9919} & 42.864 / 0.9858\\
 \bottomrule
\end{tabular}}
\end{table*}
\begin{figure*}[h!]
    \centering 
    \includegraphics[scale=0.64]{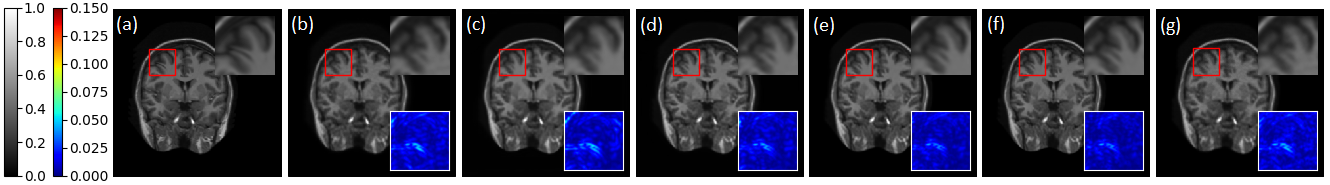}
    \caption{Qualitative results for ablation study of the proposed model. (a) GT, reconstruction results for (b) $1^{st}$, (c) $2^{nd}$, (d) $3^{rd}$, (e) $4^{th}$, (f) $5^{th}$, and (g) $6^{th}$ network settings. Inset- top right: the zoomed in region enclosed by the red box, bottom right: absolute difference between the zoomed in region and its corresponding GT. Best viewed in 200\% zoom.}
    \label{abl}
\end{figure*}
\vspace{-4mm}
\begin{table*}[h!]
\centering
\large
\caption{Quantitative comparison of various activation functions}
\label{tab_acts}
\resizebox{1.99\columnwidth}{!}{
\begin{tabular}{cccccccc} 
 \toprule
 \rule{0pt}{9pt} Activation & $\mathbb{C}$ReLU & $\mathbb{C}$PReLU & $z$ReLU & Cardioid & PP-SS & TIP-SS & PC-SS\\
 \midrule
 \rule{0pt}{9pt} PSNR (dB) / mSSIM & 45.044 / 0.9919 & 45.165 / 0.9920 & 35.991 / 0.9690 & 45.128 / 0.9919 & 45.066 / 0.9919 & 45.429 / 0.9925 & \textbf{45.678} / \textbf{0.9927}  \\
 \bottomrule
\end{tabular}}
\end{table*}
\begin{figure*}[h!]
    \centering 
    \includegraphics[scale=0.57]{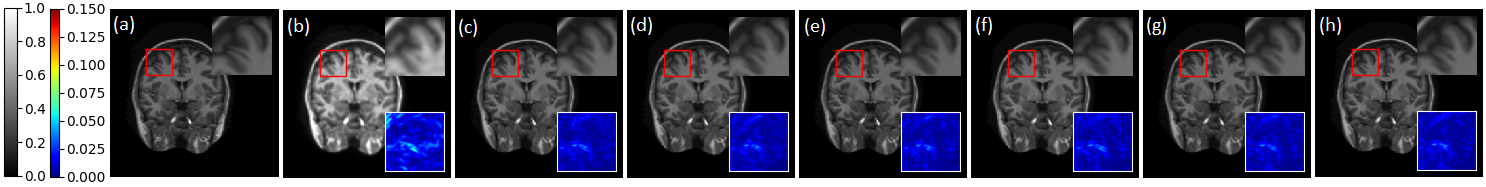}
    \caption{Qualitative comparison of various activation functions. (a) GT, reconstruction results using (b) $z$ReLU, (c) $\mathbb{C}$ReLU, (d) $\mathbb{C}$PReLU, (e) Cardioid, (f) PP-SS, (g) TIP-SS, and (h) PC-SS activation functions. Inset- top right: the zoomed in region enclosed by the red box, bottom right: absolute difference between the zoomed in region and its corresponding GT. Best viewed in 200\% zoom.}
    \label{acts}
\end{figure*}
\subsubsection{Overall Loss} The overall loss $L$ which is used to train the generator, is formulated as a weighted sum of all the losses presented above:
\begin{equation}
\label{gen_loss}
    L = \lambda_1L_{GAN}+\lambda_2L_{\ell_1}+\lambda_3L_{mSSIM}+\lambda_4L_{wvt}.
\end{equation}
In this work, $\lambda_1=0.01$, $\lambda_2=20$, $\lambda_3=1$, and $\lambda_4=100$.
\par Once the training is complete, a single forward pass through the trained generator is used to obtain the reconstructed image.
\vspace{-5mm}
\subsection{Training settings}
\par For implementing the models, Keras framework with TensorFlow backend is used. The models are trained using 4 NVIDIA GeForce GTX 1080 Ti GPUs. In this work, the batch size is set as 16. In the generator, each layer produces 32 feature maps. A total of 4 RRDBs are present in the botteleneck layer, the number of dense blocks is set as 8, and $\alpha$ is 0.2. The absolute value of the discriminator weights is clipped at 0.05, and $n_D$ is set as 3. For training the models, we use Adam optimizer \cite{adam}, with $\beta_1 = 0.5$ and $\beta_2=0.999$. The initial LR is set as $10^{-4}$, with a decay of $1.39\times10^{-3}$, so that it becomes $1/10^{th}$ of the initial value after 5 epochs.
\vspace{-2mm}
\section{Results and Discussion}
\subsection{Datasets} 
\par We evaluate our models on three datasets, namely MICCAI 2013 grand challenge dataset \cite{dataset} from which we use T$_1$ weighted MR images of brain, MRNet dataset \cite{mrnet} from which we use coronal MR images of knee, and fastMRI dataset \cite{fastmri} from which we use coronal proton density weighted single-coil MR images of knee. The fastMRI dataset contains complex-valued images, and they are cropped to the center to obtain $320\times 320$ image size. The other two datasets provide only real-valued images of size $256\times 256$. In order to improve the robustness to noise, we apply data augmentation by adding synthetic 10\% and 20\% complex Gaussian noise in the Fourier space \cite{robustgan}, before undersampling. The proportion of noise-free and images with 10\% and 20\% noise is kept same as in \cite{robustgan}, for all the three datasets. For the MICCAI 2013 dataset, 20\,787 images are used for training, which are taken from the train set of the dataset. For testing, 2000 images are selected at random from the test set provided in the dataset. For the MRNet and fastMRI datasets, 12\,500 images are randomly chosen for training, and 2000 non-overlapping images are chosen for testing. In order to observe the performance for various levels of noise, the testing is carried out for three sets: noise-free images (i.e. images without synthetic noise), images with 10\% noise, and images with 20\% noise.
\par In order to obtain the undersampled data, we use various sampling masks. For the ablation studies and comparisons, we use 30\% 1D Gaussian (1D-G) or cartesian sampling. Later, we show the performance of our approach for different sampling ratios (10\%, 20\%, 30\%) and patterns (1D-G, radial, spiral). The radial and spiral sampling patterns are generated using the Sigpy package\footnote{https://github.com/mikgroup/sigpy/tree/master/sigpy}.
\vspace{-2mm}
\subsection{Ablation Studies}
\par Table \ref{tab_abl} and Fig. \ref{abl} show the quantitative and qualitative results, respectively, for the ablation study of the model to showcase the importance of various components of the proposed model. These results are reported for 30\%  1D Gaussian (1D-G) undersampled images, from the MICCAI 2013 dataset. The first case demonstrates a real-valued GAN model which consists of a U-net based generator model without RRDBs, without the dense connections in the contracting and expanding paths, with $\mathbb{C}$ReLU activation and with $\lambda_4=0$ in (\ref{gen_loss}). In the next case, the complex-valued equivalent of the previous model is considered. In the third case, the effect of adding RRDBs in the last layer of the complex-valued U-net is observed. In the fourth case, the addition of dense connections in the complex-valued U-net is considered. In the fifth case, the effect of including $L_{wvt}$ by setting $\lambda_4=100$ is observed. Evidently, each step results in significant improvement in the quantitative metrics of peak signal-to-noise ratio (PSNR) and mSSIM, as well as in the qualitative reconstructions demonstrated in Fig. \ref{abl}. It is noteworthy that despite a $40\%$ decrease in the number of generator parameters, the complex-valued equivalent ($2^{nd}$ setting) outperforms the real-valued GAN model presented as the $1^{st}$ setting. 
\par In order to further highlight the importance of using complex-valued representations and their ability to produce high-quality reconstructions, the real-valued version of the fifth case of the ablation study is considered. In this model, each convolutional layer is converted from complex to real, but twice the number of output channels are present in each layer. The results for this model are shown in the last column of Table \ref{tab_abl}. It is observed that despite having more than twice the number of trainable parameters, this model significantly underperforms when compared to its complex-valued counterpart ($5^{th}$ setting). For the rest of the results, we use the $5^{th}$ network settings.
\vspace{-2mm}
\subsection{Activation Functions}
\par Table \ref{tab_acts} and Fig. \ref{acts} show the quantitative and qualitative results for comparing various activation functions, respectively. These results are also reported for 30\% 1D-G undersampled images, from the MICCAI 2013 dataset. It is observed that $z$ReLU, which only allows the inputs lying in the first quadrant to pass and blocks all other inputs, has the worst performance. $\mathbb{C}$PReLU outperforms $\mathbb{C}$ReLU as well as $z$ReLU. This may be because unlike these activation functions, $\mathbb{C}$PReLU does not suppress input information strongly for non-zero values of parameters $\beta_R$ and $\beta_I$. Moreover, it has an adaptable shape which also allows it to bypass issues like dead neurons, commonly observed in $\mathbb{C}$ReLU. It is observed that cardiod performs better than aforementioned activations, as it is sensitive to phase, and PP-SS obtains a similar performance as cardioid. However, these phase preserving activation functions slightly underperform as compared to $\mathbb{C}$PReLU, which does not have the phase preservation capabilities. Further, when the possibility of a negative gain is introduced in PP-SS, the resulting activation TIP-SS obtains a significantly better performance. This indicates that if phase change is allowed, the flexibility thus produced leads to a boost in performance. The same is true for the proposed PC-SS, after the parameter $\phi$ is introduced. This experiment not only highlights the superiority of PC-SS, but also shows that phase preserving activation functions prove less effective in a complex GAN framework. Although all $\mathbb{C}$PReLU, TIP-SS and PC-SS are not phase preserving activations, the superior performance of the latter two functions can be attributed to a higher flexibility in the resulting shape and sensitivity to the input phase. 

\par The rest of the results are reported by using proposed PC-SS activation, as it outperforms all the other activation functions by a considerable margin. This final model takes 37.87 ms to reconstruct each MR image. As the inference time is of the order of milliseconds, the proposed approach is viable for real-time reconstruction.
\vspace{-2mm}
\subsection{Comparison with State-of-the-art}
\subsubsection{MICCAI 2013 Dataset}  
\begin{figure}[h!]
    \centering 
    \includegraphics[scale=0.76]{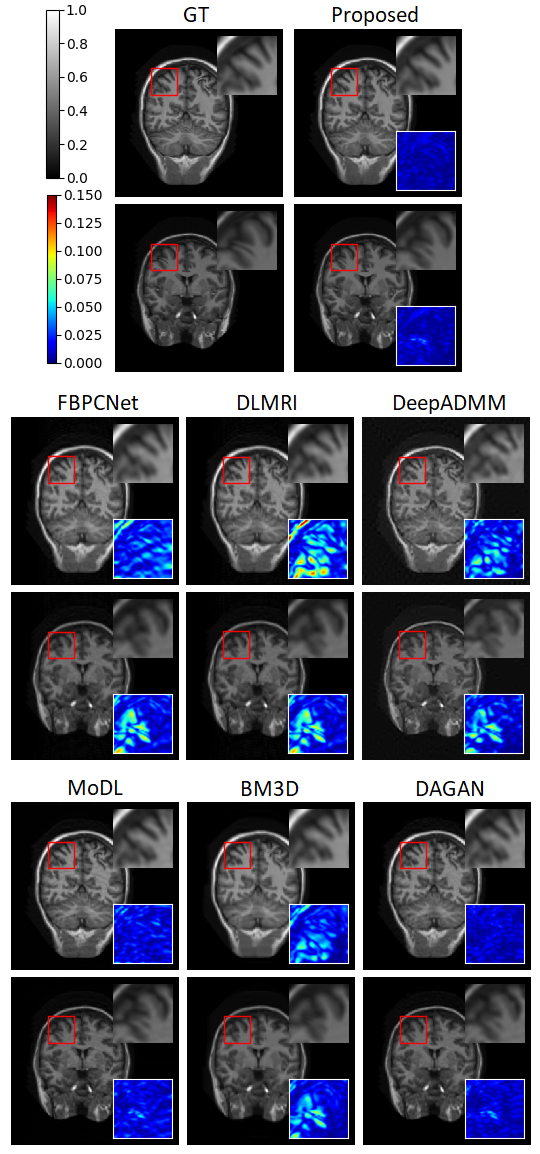}
    \caption{Qualitative results and comparison of the proposed method for two images taken from the MICCAI 2013 dataset. Inset- top right: the zoomed in region enclosed by the red box, bottom right: absolute difference between the zoomed in region and its corresponding GT.}
    \label{bigfig}
\end{figure}

 Fig. \ref{bigfig} demonstrates the qualitative results of our final model and comparison with SOTA methods for reconstruction of 30\% 1D-G undersampled images. We used the publicly available official implementations of the SOTA methods to analyze their performances on the test set. It is observed that the proposed model is able to reconstruct high-quality images by preserving most of the structural details of critical importance. As seen in the reconstructed results for the first image, some of the methods like FBPCNet, DLMRI, and BM3D are unable to properly reconstruct the high-frequency content present in the form of the edges in the zoomed in region. Also, most of the SOTA methods obtain blurry reconstruction results which might be due to the use of $l_2$ loss. This can be clearly seen by the quality of reconstruction of the fine details present in the zoomed in region of the second image. FBPCNet, DLMRI, DeepADMM, and BM3D obtain an overly smooth low-quality reconstruction, which appears to be slightly out of focus. Moreover, one can also observe the presence of subtle artefacts in the background of the zoomed in region of the first image. On close inspection of the lower half of the first image, it can be observed that FBPCNet, DLMRI, DeepADMM produce outputs with aliasing artefacts. The reconstructed images produced by DAGAN, although significantly better than other methods, do not capture subtle details. The visual representations of the proposed scheme show that it preserves the fine structural and textural details and most closely aligns with the GT, as compared to the aforementioned SOTA methods. 
\vspace{-2mm}
\begin{table}[h!]
\large
\centering
\caption{Quantitative comparison with previous methods using MICCAI 2013 dataset}
\label{comp1}
\resizebox{\columnwidth}{!}{
\begin{tabular}{cccc} 
 \toprule
 \rule{0pt}{9pt} \multirow{2}{*}{Method} & \multicolumn{1}{c}{Noise-free images} & \multicolumn{1}{c}{10\% noise level} & \multicolumn{1}{c}{20\% noise level} \\ 
 \cmidrule{2-4}
 \rule{0pt}{9pt}  & \multicolumn{3}{c}{PSNR (dB) / mSSIM} \\
 \midrule
  \rule{0pt}{9pt} FBPCNet\cite{fbpcnet} & 35.996 / 0.8655 & 34.025 / 0.6011 & 31.421 / 0.4226 \\
  \rule{0pt}{9pt} DLMRI\cite{dlmri} & 37.405 / 0.8732 & 34.144 / 0.6140 & 31.564 / 0.4346 \\
 \rule{0pt}{9pt} DeepADMM\cite{deepadmm} & 41.545 / 0.8946 & 39.078 / 0.8105 & 35.373 / 0.6000 \\
  \rule{0pt}{9pt} MoDL\cite{modl} & 42.383 / 0.9760 & 40.204 / 0.9481 & 36.844 / 0.8572 \\
  \rule{0pt}{9pt} BM3D\cite{bm3d} & 42.521 / 0.9764 & 37.836 / 0.7317 & 33.657 / 0.4947 \\
 \rule{0pt}{9pt} DAGAN\cite{dagan} & 43.329 / 0.9860 & \textbf{42.006} / 0.9814 & \textbf{39.160} / 0.9619 \\
 \rule{0pt}{9pt} Proposed & \textbf{45.678} / \textbf{0.9927} & 41.809 / \textbf{0.9838} & 39.083 / \textbf{0.9718} \\
 \bottomrule
\end{tabular}}
\end{table}

\par Table \ref{comp1} illustrates the quantitative comparison of the proposed method with the six SOTA approaches. These results are reported for 30\% 1D-G undersampled images. It can be observed  that there is an appreciable boost in both PSNR and mSSIM values obtained by the proposed approach as compared to the existing approaches. On comparing the results for images with 10\% and 20\% noise, it is observed that some of the methods like FBPCNet, DLMRI, and BM3D experience a steep decline in the reconstruction quality with the addition of noise, indicating a lack of robustness. The use of the augmentation technique with noisy images while preparing the training data for the proposed method increases its robustness towards noise. The values obtained by the proposed method for noisy images are significantly better than the other methods, but they are comparable to DAGAN. The reason for this might be the high complexity of the model, which allows superior reconstruction quality in the noise-free case at the cost of an increase in the sensitivity to the noise-level \cite{degrads}.

\begin{table}[h!]
\large
\centering
\caption{Quantitative results and comparison using MRNet dataset}
\label{comp_mrnet}
\resizebox{0.99\columnwidth}{!}{
\begin{tabular}{cccc} 
 \toprule
 
 \rule{0pt}{9pt} \multirow{2}{*}{Method} & \multicolumn{1}{c}{Noise-free images} & \multicolumn{1}{c}{10\% noise level} & \multicolumn{1}{c}{20\% noise level} \\ 
 \cmidrule{2-4}
 \rule{0pt}{9pt}  & \multicolumn{3}{c}{PSNR (dB) / mSSIM} \\
 \midrule
 \rule{0pt}{9pt} MoDL\cite{modl} & 28.471 / 0.8643 & 27.659 / 0.7740 & 25.903 / 0.6303 \\
 \rule{0pt}{9pt} DAGAN\cite{dagan} & 31.529 / 0.8754 & 30.452 / 0.8182 & 28.267 / 0.7098 \\
 \rule{0pt}{9pt} Proposed & \textbf{34.010} / \textbf{0.9306} & \textbf{33.031} / \textbf{0.9097} & \textbf{31.670} / \textbf{0.8802} \\
 \bottomrule
\end{tabular}}
\end{table}
\vspace{-2mm}
\begin{figure}[h!]
    \centering 
    \includegraphics[scale=0.67]{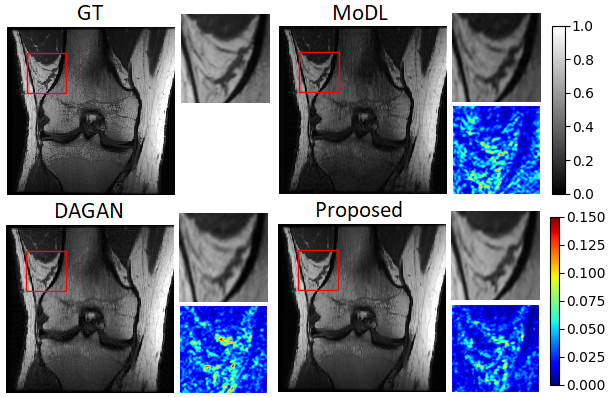}
    \caption{Qualitative results and comparison of the proposed method on images obtained from the MRNet dataset. Inset- top right: the zoomed in region enclosed by the red box, bottom right: absolute difference between the zoomed in region and its corresponding GT.}
    \label{res_mrnet}
\end{figure}
\begin{figure*}[!bh]
    \centering 
    \includegraphics[scale=0.75]{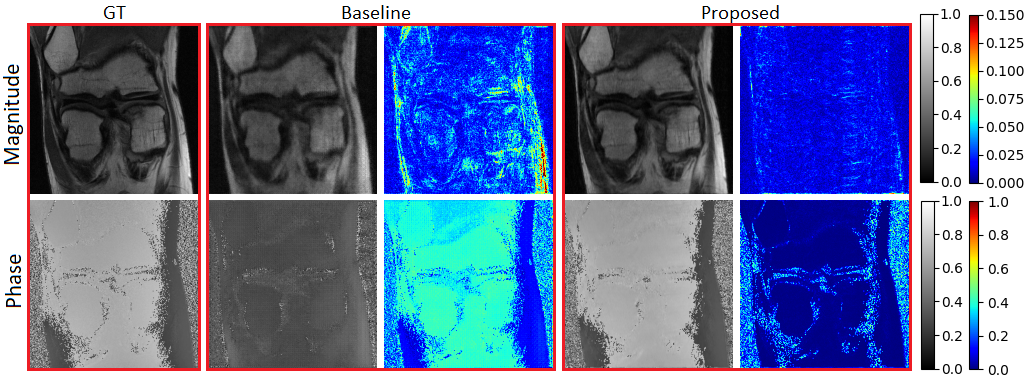}
    \caption{Qualitative results of the proposed method and comparison with the baseline for a complex-valued image from fastMRI dataset. For both the methods, the two columns show the reconstructed outputs, and their absolute difference with the GT.}
    \label{fastmri}
\end{figure*}
\subsubsection{MRNet Dataset} To further analyze the effectiveness of the proposed method compared to SOTA deep learning based methods, namely MoDL and DAGAN, we  perform another set of experiments using the MRNet dataset. Table \ref{comp_mrnet} and Fig. \ref{res_mrnet} show the quantitative and qualitative results of the proposed method using 30\% 1D-G undersampled images. It is observed that the proposed method significantly outperforms both MoDL and DAGAN for noise-free as well as noisy images. The proposed method obtains better PSNR and mSSIM results for images with 20\% noise as compared to those obtained by MoDL and DAGAN for noise-free images. The qualitative results as well the difference between the reconstructed output and the GT also demonstrate that the proposed method obtains superior reconstruction results.
\subsubsection{fastMRI dataset} 
In order to evaluate the performance of our approach on complex-valued images, we perform another set of experiments using the fastMRI dataset. For the purpose of comparison, we implement a baseline based on DAGAN. Instead of taking the magnitude images as input, the baseline takes the complex-valued images concatenated as two real-valued channels. The output is also modified in the same way. We also increase the number of trainable parameters of the U-net based generator by increasing the number of output channels in each layer. Magnitude images are used for computation of adversarial and VGG loss, while the complex-valued images are used for the MSE loss, and frequency loss \cite{dagan}. Table \ref{comp_fastmri} and Fig. \ref{fastmri} show the quantitative and qualitative results of the proposed method and the baseline using 30\% 1D-G undersampled images. 
\vspace{-2mm}
\begin{table}[h!]
\large
\centering
\caption{Quantitative results and comparison using fastMRI dataset}
\label{comp_fastmri}
\resizebox{0.99\columnwidth}{!}{
\begin{tabular}{ccccc} 
 \toprule
 \rule{0pt}{9pt} \multirow{2}{*}{Method} & \multirow{2}{*}{Params} & \multicolumn{1}{c}{Noise-free images} & \multicolumn{1}{c}{10\% noise level} & \multicolumn{1}{c}{20\% noise level} \\ 
 \cmidrule{3-5}
\rule{0pt}{9pt} & & \multicolumn{3}{c}{PSNR (dB) / mSSIM}  \\
 \midrule
 \rule{0pt}{9pt} Baseline & 133M & 29.911 / 0.5885 &  29.820 / 0.5885   &  29.576 / 0.5743  \\
 \rule{0pt}{9pt} Proposed & 1.7M & \textbf{34.593} / \textbf{0.7893} & \textbf{34.416} / \textbf{0.7870} & \textbf{33.804} / \textbf{0.7673} \\
 \bottomrule
\end{tabular}}
\end{table}

\par We see that despite having significantly fewer trainable parameters (params in Table \ref{comp_fastmri}) in the generator as compared to the baseline, the proposed approach is able to obtain superior-quality reconstruction of the complex-valued images. This is highlighted by the similarity of both the recovered magnitude and phase images to the corresponding GT. The reconstructed output by the baseline demonstrates a significant loss in both magnitude and phase information, with the former being visibly blurry and the latter showing alarmingly high differences. Moreover, we can observe that our approach obtains robust reconstructions in the presence of noise in the undersampled measurements. These observations reiterate the importance of complex-valued operations for processing complex-valued data.
\vspace{-2mm}

\begin{figure*}[!th]
    \centering 
    \includegraphics[scale=0.85]{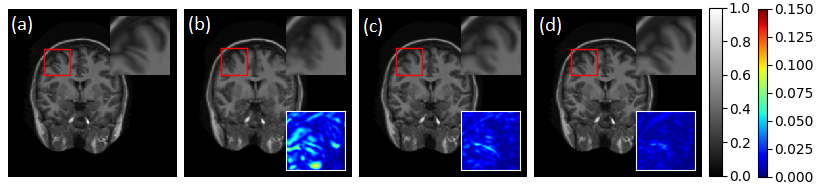}
    \caption{Qualitative results for various sampling ratios using 1D-G undersampled images from MICCAI 2013 dataset. (a) GT, reconstruction results of the proposed approach for (b) 10\%, (c) 20\%, and (d) 30\% sampling ratios. Inset- top right: the zoomed in region enclosed by the red box, bottom right: absolute difference between the zoomed in region and its corresponding GT.}
    \label{mask_ratio}
\end{figure*}
\begin{figure*}[!th]
    \centering 
    \includegraphics[scale=0.63]{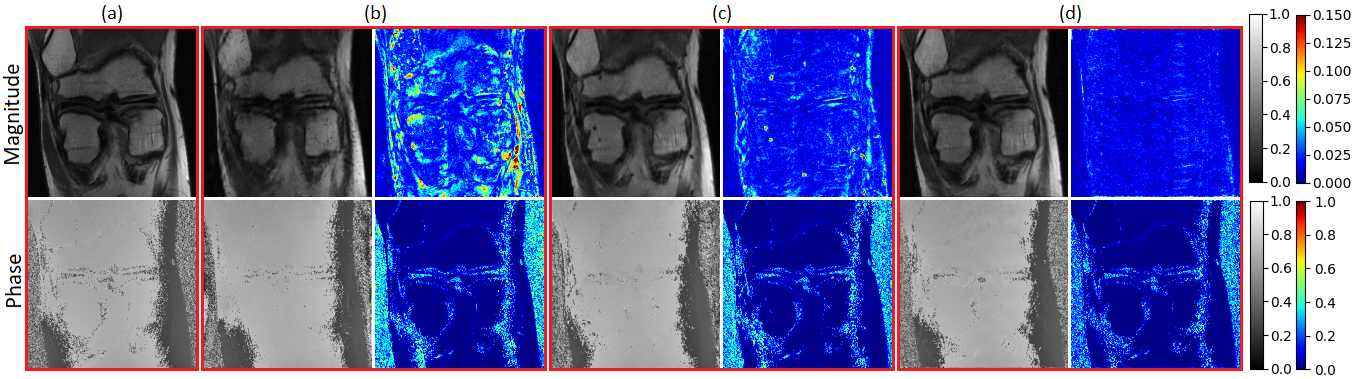}
    \caption{Qualitative results for various sampling ratios using 1D-G undersampled complex-valued images from fastMRI dataset. (a) GT, reconstruction results of the proposed approach for (b) 10\%, (c) 20\%, and (d) 30\% sampling ratios. For all the ratios, the two columns show the reconstructed outputs, and their absolute difference with the GT.}
    \label{mask_ratio_fastmri}
\end{figure*} 
\subsection{Effect of Sampling Masks}
\subsubsection{Sampling Ratio} Figs. \ref{mask_ratio} and \ref{mask_ratio_fastmri} show the qualitative results of the proposed method for various sampling ratios, for 1D-G undersampled images, taken from the MICCAI 2013 and the fastMRI dataset, respectively. It is observed that the proposed approach is able to obtain high-quality reconstructions for 20\% and 30\% undersampled images for both the datasets. Table \ref{tab_ratios} shows the quantitative results for this set of experiments. For the MICCAI 2013 dataset, the results obtained for 20\% undersampled images are quantitatively as well as qualitatively better than the results obtained by some of the SOTA methods (Table \ref{comp1}) for 30\% undersampled images. For fastMRI dataset, the results obtained for even 10\% undersampled images are better than the results obtained by the baseline (Table \ref{comp_fastmri}) for 30\% undersampled images. These observations further highlight the superior quality reconstructions achieved by the proposed approach as compared to SOTA methods. For 10\% undersampled images from both the aforementioned datasets, the proposed approach is able to obtain a de-aliased output where the contrast as well as a significant portion of structural content has been preserved.  However, it is evident that a highly faithful reconstruction may not be achieved for this ratio in both the cases. This is because, the $k$-space has been highly undersampled, and only 10\% of the data has been retained.
\vspace{-1mm}
\begin{table}[h!]
\centering
\large
\caption{Quantitative results (PSNR (dB) / mSSIM) for various sampling ratios}
\label{tab_ratios}
\resizebox{0.9\columnwidth}{!}{
\begin{tabular}{cccc} 
 \toprule
 \rule{0pt}{9pt} \multirow{2}{*}{Dataset}  & \multicolumn{3}{c}{Sampling Ratio} \\ 
 \cmidrule{2-4}
 \rule{0pt}{9pt}  & 10\% & 20\% & 30\% \\
 \midrule
 \rule{0pt}{9pt} MICCAI 2013 & 35.799 / 0.9485 & 41.396 / 0.9817 & 45.678 / 0.9927 \\
 \rule{0pt}{9pt} fastMRI & 30.596 / 0.6673 & 32.512 / 0.7356 & 34.593 / 0.7893 \\
 \bottomrule
\end{tabular}}
\end{table}
\vspace{-1mm}
\begin{figure*}[!th]
    \centering 
    \includegraphics[scale=0.85]{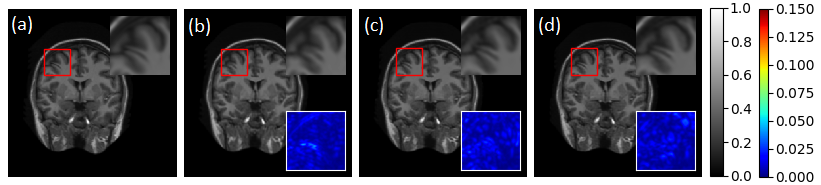}
    \caption{Qualitative results for various sampling patterns using 30\% undersampled images from MICCAI 2013 dataset. (a) GT, reconstruction results of the proposed approach for (b) 1D-G, (c) radial, and (d) spiral sampling patterns. Inset- top right: the zoomed in region enclosed by the red box, bottom right: absolute difference between the zoomed in region and its corresponding GT.}
    \label{mask_pat}
\end{figure*}
\begin{figure*}[!th]
    \centering 
    \includegraphics[scale=0.63]{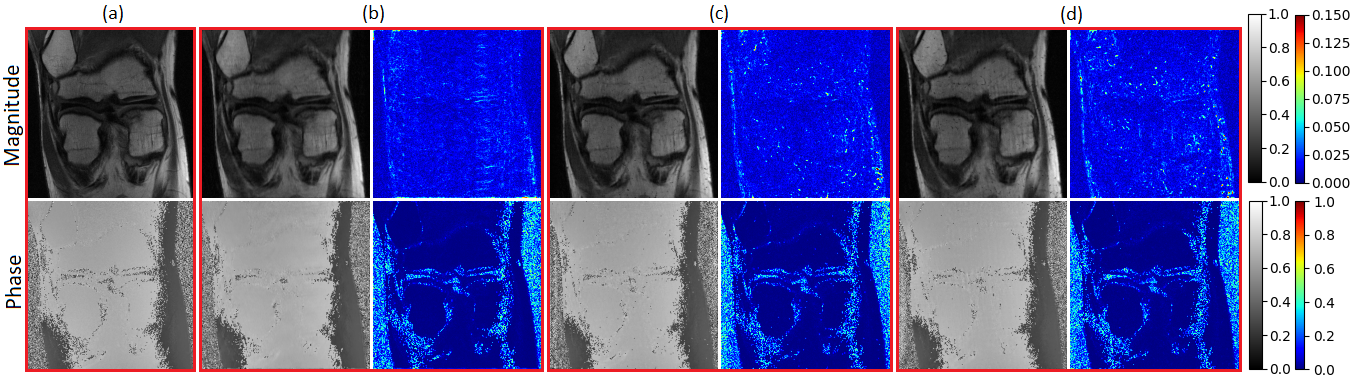}
    \caption{Qualitative results for various sampling patterns using 30\% undersampled complex-valued images from fastMRI dataset. (a) GT, reconstruction results of the proposed approach for (b) 1D-G, (c) radial, and (d) spiral sampling patterns. For all the patterns, the two columns show the reconstructed outputs, and their absolute difference with the GT.}
    \label{mask_pat_fastmri}
\end{figure*}
\subsubsection{Sampling Pattern} Figs. \ref{mask_pat} and \ref{mask_pat_fastmri} demonstrate the visual outputs generated by the proposed approach for comparing various sampling patterns, using 30\% undersampled images, from the MICCAI 2013 dataset and the fastMRI dataset, respectively. For both the datasets, highly accurate recoveries are achieved for all three sampling masks, as evident from the difference between the generated output and the GT. For the MICCAI 2013 dataset, this difference is close to zero, signifying the proposed method's ability to reconstruct the finest details as seen in the zoomed in region. For the fastMRI dataset, the proposed method obtains high-quality reconstruction of both magnitude and phase images. These experiments show that our method generalizes well to various sampling patterns. This is also supported by the quantitative results for this set of experiments shown in Table \ref{tab_patterns}.
\vspace{-2mm}
\begin{table}[h!]
\centering
\large
\caption{Quantitative results (PSNR (dB) / mSSIM) for various sampling patterns}
\label{tab_patterns}
\resizebox{0.9\columnwidth}{!}{
\begin{tabular}{cccc} 
 \toprule
  \rule{0pt}{9pt} \multirow{2}{*}{Dataset}  & \multicolumn{3}{c}{Sampling Pattern} \\ 
 \cmidrule{2-4}
 \rule{0pt}{9pt}  & 1D-G & Radial & Spiral \\
 \midrule
 \rule{0pt}{9pt} MICCAI 2013 & 45.678 / 0.9927 & 46.629 / 0.9922 & 46.747 / 0.9929 \\
 \rule{0pt}{9pt} fastMRI & 34.593 / 0.7893 & 34.173 / 0.7819 & 33.918 / 0.7745 \\
 \bottomrule
\end{tabular}}
\end{table}
\vspace{-3mm}
\subsection{Zero-shot Inference}
\par
\begin{figure}[h!]
    \centering 
    \includegraphics[scale=0.75]{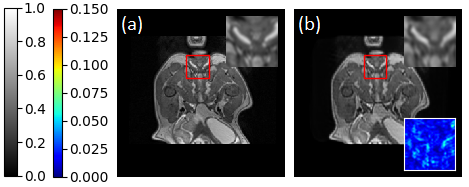}
    \caption{Qualitative results of zero-shot inference experiment. (a) GT, (b) reconstruction results of the proposed model trained on images from MICCAI 2013 dataset. Inset- top right: the zoomed in region enclosed by the red box, bottom right: absolute difference between the zoomed in region and its corresponding GT.}
    \label{zs}
\end{figure}

 In this experiment, the model trained on 30\% 1D-G undersampled brain images from the MICCAI 2013 dataset is tested for reconstruction of 30\% 1D-G undersampled images of canine legs from the MICCAI 2013 challenge. This model achieves an average PSNR of 42.949 dB and mSSIM of 0.9864, when inferred for 2000 test images. 
 The qualitative results of this zero-shot inference are presented in Fig. \ref{zs}. This shows that although the images of canine legs (used to test the model) are of a completely different anatomy as compared to those used for training, our approach is able to obtain high-quality reconstruction. 
 \par The proposed scheme using the $L_{\ell_{1}}$, $L_{mSSIM}$, $L_{wvt}$ loss functions during training tries to align the reconstructed MR image closely to the GT, in effect trying to reduce the possibility of hallucination by GAN. All the aforementioned experiments, including the zero-shot inference, help demonstrate that the proposed framework has displayed no signs of hallucination. An expert radiologist confirmed the same, after examining the GT and reconstructed images.
\vspace{-3mm}
\section{Conclusion}
In this paper, we proposed a novel Co-VeGAN framework to combat the fundamental weakness of SOTA deep learning based frameworks for CS-MRI reconstruction, namely their inability to process the inherently complex-valued MR images. We also propose a learnable complex-valued activation function PC-SS, which is sensitive to the variations in the input phase. Detailed analyses have shown that the proposed method obtains high-quality reconstructions for both real and complex-valued datasets, which opens up the possibility of its use for other medical imaging applications reliant on complex-valued data. The fast inference step allows for a real-time implementation of the method, while the less number of generator parameters considerably reduces the storage requirement. 
\section*{Acknowledgement}
The authors would like to thank Dr. Vijinder Arora, Professor and Head, Department of Radiology and Radio Diagnosis, Sri Guru Ram Das Institute of Medical Sciences and Research, and senior radiologist at Nijjar Scan and Diagnostic centre, Amritsar, Punjab, for her expert opinion regarding the comparison of GT and reconstructed MR images.
\bibliography{name/refs.bib}{}
\bibliographystyle{ieeetr}

\end{document}